\documentclass[traditabstract,rnote]{aa}

\usepackage{natbib}
\bibpunct{(}{)}{;}{a}{}{,}
\usepackage{txfonts}
\usepackage{graphicx}
\usepackage{epstopdf}

\title{When two become one:\\ an apparent QSO pair turns out to be a single quasar}
\titlerunning{An apparent pair turns out to be a single quasar}

\author{
Guido Cupani\inst{\ref{inst:OATs}}\thanks{E-mail address:\texttt{cupani@oats.inaf.it}.}
\and Stefano Cristiani\inst{\ref{inst:OATs}}
\and Valentina D'Odorico\inst{\ref{inst:OATs}}
\and Bo Milvang-Jensen\inst{\ref{inst:2}}
\and Jens-Kristian Krogager\inst{\ref{inst:2}}
}

\institute{
INAF -- Osservatorio Astronomico di Trieste, Via Tiepolo 11, 34131 Trieste, Italy\label{inst:OATs}
\and Dark Cosmology Centre, Niels Bohr Institute, University of Copenhagen, Juliane Maries Vej 30, 2100 Copenhagen {\O}, Denmark\label{inst:2}}

\date{Received / Accepted }

\begin{document}

\abstract{
We report on our serendipitous discovery that the objects Q 01323-4037 and Q 0132-4037, listed in the V\'eron-Cetty \& V\'eron catalog (2006) as two different quasars, are actually a quasar and a star. We briefly discuss the origin of the misidentification, and provide a refined measurement of the quasar redshift.
}


\maketitle

\section{Introduction}

\object{Q 01323-4037} and \object{Q 0132-4037} are listed as two separate quasi-stellar objects (QSOs) in the last editions of the \emph{catalogue of quasars and active nuclei} \citep{VeronCetty:2003p565,VeronCetty:2006p231,VeronCetty:2010p564}. According to this catalog, the two objects (A and B for simplicity) have similar redshift, $z_\mathrm A=2.100$, $z_\mathrm B=2.150$. Their angular separation in the sky, as computed from the J2000 coordinates ($\alpha_\mathrm A=01$ $34$ $32.5$, $\delta_\mathrm A=-40$ $22$ $08$ and $\alpha_\mathrm B=01$ $34$ $32.2$, $\delta_\mathrm B=-40$ $21$ $33$, respectively), is approximately $35$ arcsec, making the seeming QSO pair a suitable candidate for a tomographic study of the inter-galactic medium along close lines of sight.

\begin{figure}[htbp]
\begin{center}
\resizebox{\hsize}{!}{\includegraphics{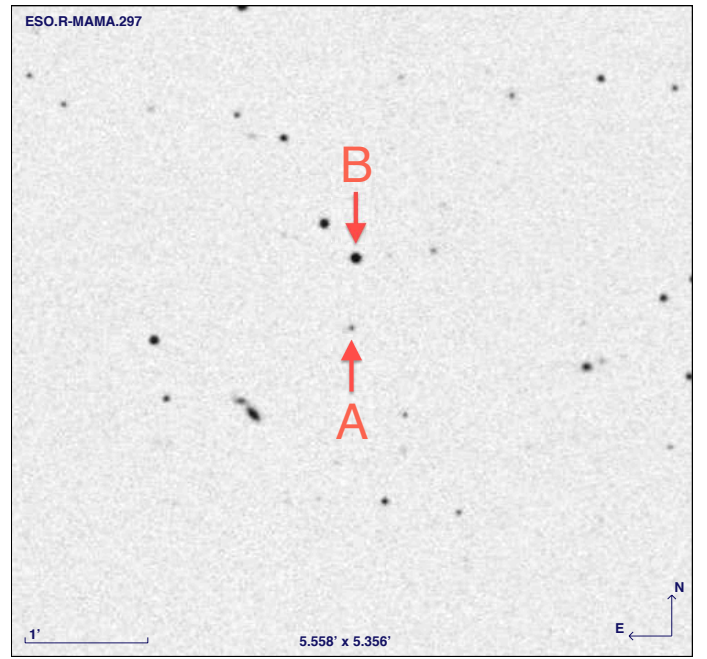}}
\caption{Finding chart with the position of object A (a QSO) and object B (a star mistakenly identified as a separate QSO). The image is centered on object A (J2000 coordinates $\alpha_\mathrm A=01$ $34$ $32.5$, $\delta_\mathrm A=-40$ $22$ $08$).}
\label{fig:chart}
\end{center}
\end{figure}

We observed the two objects in November, 2010 with the single target, medium resolution spectrograph X-shooter \citep{DOdorico:2006p815} at the \emph{Very Large Telescope} (VLT), in the context of a GTO program. Fig.~\ref{fig:chart} reproduces the finding chart used for our observation, with the position of A and B according to \citet{VeronCetty:2010p564}. We extracted the 1D spectra of the objects for the three X-shooter arms (UVB, VIS, and NIR) using the release 1.2.0 of the X-shooter reduction pipeline \citep{Goldoni:2006p199} and the ESO-MIDAS package. Fig.~2 shows the spectra obtained after flat-field correction, bias and sky subtraction, and flux calibration. Much to our surprise, the observation revealed that B is not a QSO, unlike A, which displays a typical QSO spectrum with a strong Lyman $\alpha$ emission at $\lambda\simeq 3820$ \AA. Apparently, a spurious object has been included in the V\'eron-Cetty \& V\'eron catalog by mistake.

\begin{figure*}[htbp]
\begin{center}
\resizebox{\hsize}{!}{\includegraphics{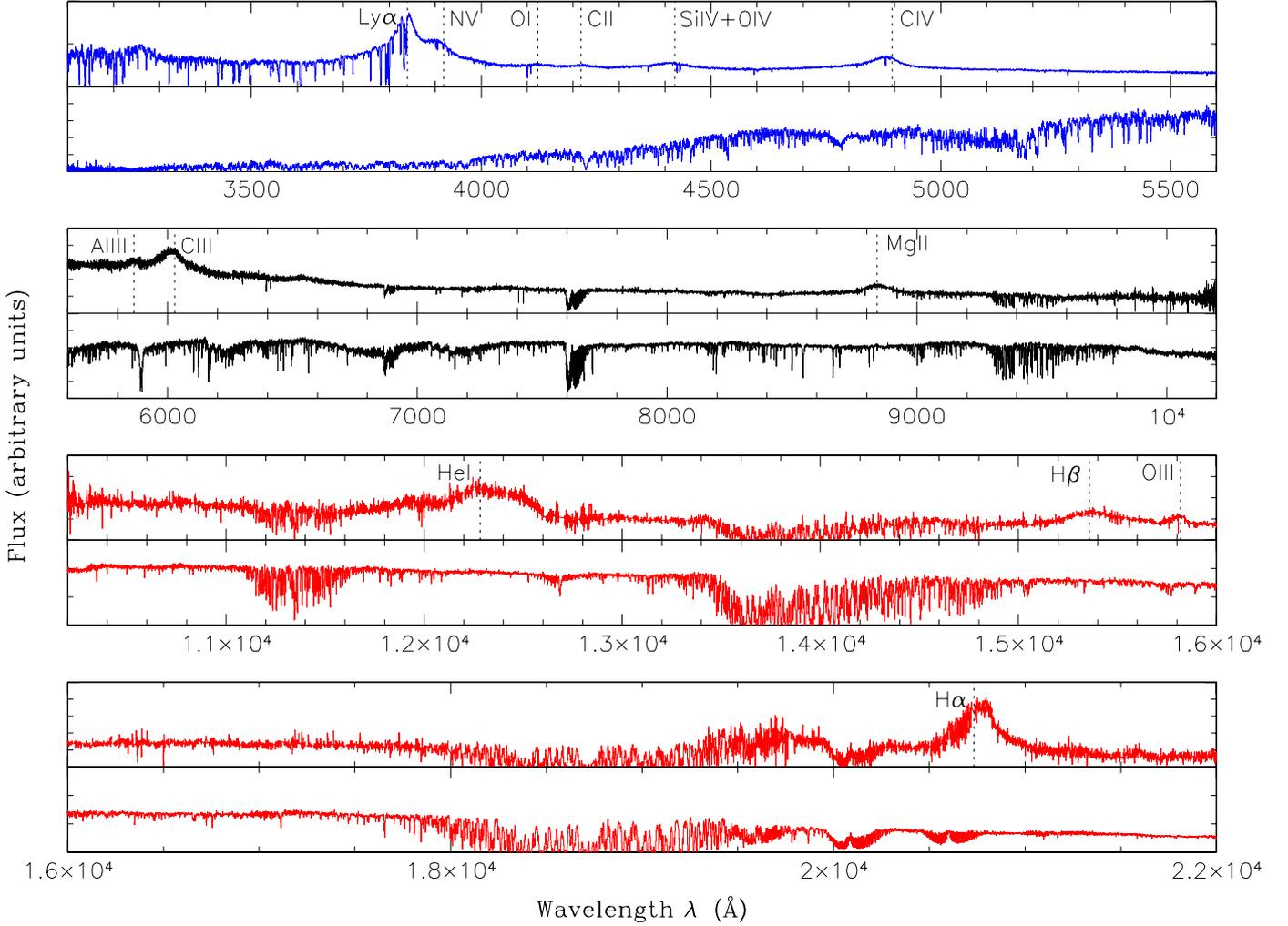}}
\caption{Spectra of object A (upper panels) and object B (lower panels) obtained with X-shooter. Flat-field correction, bias and sky subtraction, and flux calibration were performed on both spectra. Vertical dotted lines highlight the main emission features which allow to identify object A as a QSO. Similar features are totally absent in the spectrum of object B. [Color legend -- Blue: UVB arm; black: VIS arm; red: NIR arm.]}
\label{fig:spectra}
\end{center}
\end{figure*}

\section{Origin of the misidentification}

We argue that the misidentification of object B as a QSO is a consequence of poorly-constrained measurements on the object A, which is in fact the only QSO in a radius of $10$ arcmin. Different values of position and redshift attributed to A during the years were interpreted as the presence of \emph{two} QSOs in this region, one of which was incorrectly identified with B.

QSO A was first observed by \citet{Hoag:1977p647} at B1950 coordinates $\alpha=01$ $32$ $22.2$, $\delta=-40$ $37$ $18$. \citet{Osmer:1980p581} published the redshift of this objects, $z=2.15\pm 0.01$.\footnote{Quite interestingly, \citet{Hoag:1977p647} and \citet{Osmer:1980p581} are quoted in the NASA-IPAC Extragalactic Database as references for A, and in the SIMBAD Astronomical Database as references for B. This is a hint that some mismatch occurred in the identification of the two objects.} The QSO is included in the \emph{Automatic Plate Measurement} (APM) galaxy survey \citep{Maddox:1990p587} and in the catalog by \citet{Hewitt:1993p390} (labeled as B 0132-406). It is also listed in the catalog by \citet{Iovino:1996p605}, with B1950 coordinates $\alpha=01$ $32$ $21.2$, $\delta=-40$ $37$ $29$ and redshift $z=2.10$. Despite the slight differences, there is no doubt that the object is the same.

Unfortunately, the finding chart published by \citet{Hoag:1977p647} points to object B, not to object A, as the target QSO. This error (together with the poor accuracy of the published QSO position) is the most probable cause of the misidentification, which can be traced back to \citet{VeronCetty:1996p645}. Here the B1950 coordinates of the QSO appear to be $\alpha=01$ $32$ $21.12$, $\delta=-40$ $36$ $53.7$ (corresponding to J2000 coordinates $\alpha=01$ $34$ $32.27$, $\delta=-40$ $21$ $33.6$), quite different than the values previously measured, and coincident with the position of object B. The new coordinates appear in all editions of the V\'eron-Cetty \& V\'eron catalog until 2001 \citep{VeronCetty:1996p573,VeronCetty:1998p654,VeronCetty:2000p651,VeronCetty:2001p676}. During this period, all references to Q 0132-4037 \citep{Barkhouse:2001p656,Cutri:2003p658} point to object B instead of object A.  

The real QSO reappears in \citet{VeronCetty:2003p565} as an additional entry Q 01323-4037, distinct from Q 0132-4037. The authors cite \citet{Iovino:1996p605} as a reference for the first one, and \citet{Osmer:1980p581} as a reference for the second one. The slight difference in position and redshift may have led them to believe there were two different QSO, even though both \citet{Osmer:1980p581} and \citet{Iovino:1996p605} had observed the same object. The error has been reproduced in the subsequent of the catalog until today \citep{VeronCetty:2006p231,VeronCetty:2010p564}.

To summarize: the two entries Q 01323-4037 and Q 0132-4037 in the last editions of V\'eron-Cetty \& V\'eron catalog correspond in fact to a single object, QSO A, with J2000 coordinates $\alpha=01$ $34$ $32.5$, $\delta=-40$ $22$ $08$ and redshift $z$ between $2.10$ and $2.15$. Object B, whose coordinates are associated with Q 0132-4037, is not a QSO and should not be considered as such. The spectrum of object B is that of a star with spectral type early M, as seen from weak (but clear) TiO absorption bands around 620 nm and 710 nm. The spectrum shows a good resemblance to the spectrum of the M1 III star \citep{Bagnulo:2003p887}

Hereafter, we suggest to address to object A as Q 0132-4037, and to stop using the ambiguous identifier Q 01323-4037 introduced by \citet{VeronCetty:2003p565}.

\section{Refined measurement of Q 0132-4037 redshift}

According to several studies \citep{Gaskell:1982p852,Vrtilek:1985p867,Hutchings:1987p872}, redshift estimated from forbidden lines are in agreement to $\sim$$100$ km s$^{-1}$ with the systemic redshifts of QSOs as determined by stellar absorption and H \textsc{i} 21 cm emission in the host galaxies. We estimated the redshift of Q 0132-4037 by a gaussian fitting of the narrow line [O \textsc{iii}] $\lambda 5007$, which appears unblended and quite prominent in the NIR part of the spectrum (fig.~\ref{fig:spectra}). We obtained $z_{\scriptsize{\textrm{[O \tiny{\textsc{iii}}]}}}=2.1568\pm 0.0081$. The same procedure was performed for the low-ionization line Mg \textsc{ii} $\lambda 2798$ and the H$\beta$ line, giving $z_{\scriptsize{\textrm{Mg \tiny{\textsc{ii}}}}}=2.1614\pm 0.0155$ and $z_{\mathrm{H}\beta}=2.1629\pm 0.0200$, respectively. The three estimates are in agreement within the uncertainty. The combined fiducial value of the systemic redshift is estimated as $z_\mathrm{sys}=2.1583\pm 0.0067$, confirming the value by \citet{Osmer:1980p581} and ruling out the value $z=2.10$ published by \citet{Iovino:1996p605}.

\begin{acknowledgements}
Based on observations collected at the European Southern Observatory, Chile, as part of program 086.A-0076. The Dark Cosmology Centre is funded by the Danish National Research Foundation. G.~C.~would like to thank Carlo Morossi for his help in recognizing the spectral type of object B.
\end{acknowledgements}

\bibliographystyle{aa}
\bibliography{cupani-et-al_2011}

\end{document}